\documentstyle[aas2pp4]{article}          

\begin{document}

\title{Ultra-High-Density Molecular Core and Warped Nuclear Disk in the Deep
Potential of Radio-Lobe Galaxy NGC 3079}
\author{Y. Sofue$^1$,
J. Koda$^1$,
K. Kohno$^2$,
S. K. Okumura$^2$,
M. Honma$^3$,
A. Kawamura$^1$, \\ \&
Judith A. Irwin$^4$}

\affil{1. Institute of Astronomy, University of Tokyo, Mitaka,
Tokyo 181-0015,  Japan (YS: sofue@ioa.s.u-tokyo.ac.jp)\\
2. Nobeyama Radio Observatory, National Astronomical Observatory, Mitaka,
Tokyo 181-8588, Japan\\
3.  VERA Project Office, National Astronomical Observatory, Mitaka, Tokyo \\
 181-8588, and Mizusawa Astrogeodynamics Observatory, NAO, Mizusawa 023-0861,
Japan\\
4. Department of Physics, Queen's University, Kingston, Ontario, K7L 3N6, Canada  }

\def\v{\vskip 2mm}
\def\r{\hangindent=1pc  \noindent}
\def\onezero{$^{12}{\rm CO}(J$=1-0) }
\def\vlsr{$ V_{\rm LSR}$ }
\def\lco{$L_{\rm CO}$}
\def\Ico{I_{\rm CO}}
\def\Msun{M_{\odot \hskip-5.2pt \bullet}}
\def\msun{$M_{\odot \hskip-5.2pt \bullet}$}
\def\kms{km s$^{-1}$}
\def\Deg{^\circ}
\def\deg{$^\circ$}
\def\sec{$''$}
\def\htwo{ H$_2$ }
\def\ha{ H$\alpha$ }

\begin{abstract}

We have performed high-resolution synthesis observations of
the $^{12}$CO (J=1-0) line emission from the radio lobe edge-on spiral
NGC 3079 using a 7-element mm-wave interferometer at Nobeyama, which consisted
of the 45-m telescope and 6-element Array.
The molecular nuclear disk (NMD) of 750 pc radius is found to be
inclined by 20\deg\ from the optical disk, and the NMD has spiral arms.
An ultra-high-density molecular core (UHC) was found at the nucleus.
The gaseous mass of the UHC within 125 pc radius is as large as
$\sim 3\times 10^8\Msun$, an order of magnitude more massive than
that in the same area of the Galactic Center, and the mean density is
as high as $\sim 3\times10^3 {\rm ~H_2~cm^{-3}}$.
A position-velocity diagram along the major axis indicates that the rotation
curve starts at a finite velocity exceeding 300 \kms\ already from the nucleus.
The surface mass density in the central region is estimated to be as high
as $\sim 10^5 \Msun~{\rm pc}^{-2}$, producing a very deep
gravitational potential.
We argue that the very large differential rotation in such a deep
potential will keep the UHC gravitationally stable from current star formation.

Subject headings:
ISM: molecules ---
galaxies: individual (NGC 3079) ---
galaxies: ISM ---
galaxies: kinematics and dynamics ---
galaxies: nuclei  ---
galaxies: structure ---
radio lines: ISM
\end{abstract}

\section{Introduction}

NGC3079 is an edge-on galaxy known for its pronounced radio, \ha\ and X-ray
lobes emerging from the nucleus
(Hummel et al 1983; Duric et al 1983; Lord et al 1986;
Veilleux et al 1994; Pietsch et al 1999).
The nuclear outflow may originate with accretion of dense gas onto a
compact core (Irwin and Seaquist 1988), and the nucleus exhibits
LINER and Seyfert 2 nuclear activity (Lord et al 1986).
NGC 3079 has been studied in various radio wavebands in
continuum, HI and CO lines as well as maser lines
(Irwin and Seaquist 1991;   Ford et al. 1986; Duric et al 1983;
Duric and Seaquist 1988; Irwin et al 1987; Irwin et al 1988;
Young et al. 1988; Sofue and Irwin 1992; Irwin and Sofue 1992;
Sawada-Sato et al. 2000).
The galaxy's distance is 15.6 Mpc ($H_0=75$ \kms/Mpc), and the
major axis position angle of the optical disk is $167\Deg$
and the inclination angle 85\deg.

Earlier CO observations with the Nobeyama mm Array (NMA)
inferred the existence of a dense nuclear molecular disk, rotating at
a high velocity within the central 10$''$ (750 pc)
(Sofue and Irwin 1992; Irwin and Sofue 1992).
NGC3079 is a galaxy whose unusual nuclear activity can be probed by
observing the affected interstellar medium, and can be compared with
the Milky Way for its similar edge-on orientation.
Knowledge of the kinematics, morphology and density of the interstellar
medium in the immediate vicinity of nucleus is particularly important, since
the ISM can both fuel the nucleus and trigger activity.
The nuclear disk would also play a role in collimating outflows.
For this, higher resolution CO observations have been required to
determine how the morphology and kinematics of the nuclear molecular
disk are related to the nuclear activity.

In this paper we report the result of high-resolution CO-line observations
of NGC 3079 using a 7-element millimeter wave interferometer at Nobeyama
by combining the NMA (Nobeyama mm Array) and 45-m telescope,
whose code name was "RAINBOW".

\section{Observations}

The  $^{12}$CO (J=1-0) observations of NGC3079 were made
on 2000 January 18 and February 2 in the 7-element "RAINBOW"
mode at Nobeyama, which consisted of
six 10-m antennae array in AB configuration linked with the 45-m telescope.
Observations in the C and D-configuration observations
were made on 2000 March 12 and April 12, respectively.
The $UV$ coverage of the observations was ideal for the high declination.
After obtaining UV data for individual array configurations, they were
combined, and data analysis was done by using the AIPS standard packages.
The UV data were CLEANed and Fourier transformed to three dimensional
cubes in RA, Decl, and frequency space, and further
transposed  to  (RA, Decl, \vlsr) cubes.
The synthesized HPBW  was  $1''.62\times 1''.34$,
slightly elongated in the direction of position angle 110\deg,
for combined UV data of the 7-element synthesis in AB, C and D configurations
(RABCD configuration).
The beam for the 7-element AB configuration without C and D
was $1''.27 \times 1''.09$ elongated at a position angle 106\deg
(RAB configuration).
At a distance of 15.6 Mpc $1''$  corresponds to  $75$ pc.

The center position of the galaxy was taken at the radio nucleus at
RA = 09h 58m 35.02s and Dec = $55\Deg 55' 15.4''$ (epoch 1950) and the
systemic LSR velocity was taken as 1113 \kms.
The phase and band-pass calibrations were made by observing the nearby radio
source QSO 0957+561, which was measured to have a flux density of 0.68 Jy
at the observing frequency for RAB array in January and February,
and 0.58 Jy for C and D arrays in March and April 2000.
The field of view for the 7-elelment array
including the 45-m telescope (RABCD) was $20''$, and that for the 6-element
ABCD-configuration array without 45-m telescope was 60$''$.
This yielded higher sensitivity in the central 20$''$ region
than in the surrounding region.

Fig. 1 shows the integrated intensity map of the CO line from the
RABCD configuration, and a mean-velocity field
at a resolution of $1''.62 \times 1''.36$ is shown as the inset.
The major structures discussed below are indicated
by the arrows and dashed illustrations.
A DSS $B$-band image is inset in the upper-left corner.
Fig. 2 shows an integrated intensity for the central region from the
RAB configuration with a higher resolution of $1''.27\times1''.09$.
Fig. 3 shows  position-velocity (PV) diagrams
along the major axis with a $5''$ slit width by RABCD array (upper panel),
and for the innermost region with $2''$ slit width by RAB array
(lower panel).

---- Fig. 1, 2, 3 ----

\section{Molecular Structures}

\subsection{Ultra-High-Density Molecular Core}

The most remarkable feature in the intensity maps (Fig. 1 and 2) is a very
compact, intense CO concentration at the nucleus.
This "ultra-high-density core (UHC)" of molecular gas
is elongated in the north-south direction at PA=176\deg, inclined
by 9\deg\ from the main disk.
The full width of half maximum of the UHC is measured to be
$3''.3 \times 1''.8 ~ (250\times180$ pc), and therefore,
the radius is about 125 pc.

The integrated CO intensity at the nucleus is
117 Jy/beam \kms, or $\Ico = 5.0\times 10^3$ K \kms.
The H$_2$ column density toward the nucleus  is estimated to be
$N_{\rm H_2} \sim 5.0 \times 10^{23}$ H$_2$ cm$^{-2}$,
where we took a conversion factor for the centers of galaxies of
$X=1.0 \times 10^{20} {\rm H_2 ~ K^{-1} [km~s^{-1}]^{-1} }$
(Arimoto et al 1996).

The total molecular mass in the UHC is estimated to be
$M_{\rm gas}=  3\times 10^8 \Msun$.
Since the vertical direction is not resolved, we here assume that the
thickness of the UHC is of the same order of 30 pc as the molecular ring
in the Milky Way center (Sofue 1995).
Then, the mean density of molecular hydrogen is estimated to be of
the order of or higher than $\sim 3\times 10^3 {\rm H_2 cm^{-3}}$.
Hence, the UHC consists of a pile of molecular gas equivalent to
a thousand medium-sized GMCs within a 125 pc radius.

The velocity field in Fig. 1 (lower-right inset)
shows that the UHC is rotating regularly, but very rapidly.
The PV diagrams (Fig. 3) show that the rotation velocity in the
central $1''$ region rises extremely steeply, or more likely,
the velocity starts at a finite velocity of $\sim 300$ \kms\
already at the nucleus.
We may estimate the dynamical mass from the rotation velocity,
because the pressure term due to velocity dispersion is negligible
for population I gases, particularly for molecular gas.
The dynamical mass within $R=125$ pc is, then, estimated as
$M_{\rm dyn}=V_{\rm rot}^2 R/G \sim 2\times 10^9 \Msun$.
Hence, the gaseous mass in the molecular core shares about 15\% of the
dynamical mass.

\subsection{Warped Nuclear Disk with  Spiral Arms}

The molecular core is surrounded by the nuclear molecular disk (NMD) of radius
$10''$ (750 pc) elongated  in the direction of the optical major axis
(dashed ellipse in Fig. 1).
The projected minor-to-major axial ratio of the ellipse is as large as $0.40$.
Since the disk thickness is considered to be
sufficiently small, such a fat shape indicates that
the inclination angle of the disk is either 66\deg.
This implies that the NMD is not in the same plane as the outer optical
disk whose inclination angle is 85\deg.
The NMD is, therefore, warped by about 20\deg\ from the main optical disk.

The disk is superposed by two symmetrical spiral arms,
one arm extending from the north-west of the UHC
toward the north, and the other from south-east to the south, which are
illustrated by dashed lines in the inset in Fig. 1.
If we assume that the spiral arms are trailing,
the eastern half of the disk is in the
far side, and western half is near side.
This configuration is in the same sense as that of the outer optical disk.

The velocity field shows that the circular rotation is dominant.
The slightly distorted velocity field is superposed, which suggests small
non-circular motion due to spiral arms.
Since the intensity map shows spiral arms, the latter may be more likely.
In the PV diagram the spiral arms show up as inclined ridges in
the central $10''$, crossing the bright ridge of the UHC.

The total mass of molecular gas of the NMD including the outskirt
extending to radius 900 pc is estimated to be $2\times10^9\Msun$ for
the same conversion factor as above.
This shares  about 14\% of the dynamical mass within the same radius,
$1.4\times10^{10}\Msun$, for a rotation velocity of 260 \kms\ at
900 pc radius.

\subsection{Main Disk}

Fig. 1 indicates the existence of a more extended, narrower component
extended in the direction of the optical major
axis for more than $\pm 30'' ~(\pm2.2$ kpc).
This feature is  attributed to the outer main disk visible
in optical images associated with dust lanes.
This disk component is slightly lopsided in the sense that the
southern part is brighter.

The disk shows up also in the PV diagram (Fig. 3)
as the two symmetrical tilted ridges crossing the UHC and NMD.
Such PV ridges are typical for grand-designed two armed density-wave spirals.
These PV ridges are bifurcated from the maximum-velocity parts of
the PV ridges due to the NMD.
Hence, these disk arms may be continued to the spiral arms in the NMD.

\section{Rotation Curve and Mass Distribution}

Applying the envelope-tracing method (Sofue 1996)
to the PV diagrams, we have derived a rotation curve, and combined it
with an outer rotation curve obtained from HI data (Irwin and Seaquist 1991).
The obtained rotation curve for the entire galaxy is shown in Fig. 4.
The rotation curve starts already from a very high finite value in the
nucleus, followed by a usual curve in the disk and halo.
High velocities in the center, followed by a flat rotation
in the disk and halo, exhibiting a broad maximum in the
disk, are commonly found in spiral galaxies observed with sufficiently high
resolutions (Sofue et al. 1999).

Since the circular motion is dominant (Fig. 1),
we, then, use the rotation curve to directly calculate
the distribution of surface-mass density (SMD),
applying the method developed by Takamiya and Sofue (1999).
Fig. 5 shows the obtained SMD, where the thick line shows the result
for a flat-disk assumption, and the thin line for a spherical assumption
[see Takamiya and Sofue (1999) for detaileds].
The true SMD value is considered to be sandwitched by the
two lines (thick and thin) within an error of about 30\%,
except for the outermost region, where the edge effect becomes not negligible.
The SMD between $R\sim 1$ and 8 kpc is well fitted by an
exponential disk of a scale ($e$-folding) radius of 3.5 kpc.
The bulge component between 0.3 and 1 kpc can be approximated by an
$\sim e^{-(R/200 {\rm pc})^{1/4}}$ law.
The innermost region within $R\sim300$ pc shows a much steeper increase
toward the nucleus, exhibiting  an extremely high-density dynamical
core with SMD as high as $10^5 \Msun~{\rm pc}^{-2}$ at $R\sim100$ pc.

---- Fig. 4, 5 -----

\section{Discussion}

The central $R\sim 100$ pc region of NGC 3079 is filled with the UHC
with a molecular mass as high as $3\times 10^8 \Msun$.
The UHC is rotating regularly at high speed of $ \sim 300 $ \kms\ in the deep
gravitational potential.
The molecular gas density and mass are an order of magnitude greater than those
in the Galactic Center.
We stress that such an extremely high density, massive interstellar
condition was found for the first time.

We here consider the reason why such a dense core could have survived
in the gas phase, without suffering from current star formation.
The Jeans time in the UHC is of the order of $t_{\rm J} \sim 10^6$ yr.
On the other hand, the dynamical time scale for a cloud to be torn off
by the Coriolis force and differential rotation is
$ t_{\rm D} \sim dV/dR - \omega \sim 2.5\times 10^5$ yr.
Here,  $V,~ R$ and $\omega$ are the rotation velocity, radius,
and angular velocity.
Hence, $t_{\rm J} > t_{\rm D}$, so that clouds  cannot collapse to form
stars, being are kept gravitationally stable and stretched azimuthally
along the orbits by the differential rotation.
Velocity dispersion would be an alternative cause
for surpression of star formation.
However, the random motion and turbulence in molecular gas are known to be
several \kms\ or less, which will less affect the stability
compared to the effect of the high differential rotation.

Formation of the UHC will be understood as due to accretion by
non-circular streaming motion from the NMD, as readily suggested
from the spiral pattern in the NMD.
The existence of a compact, massive central mass
would cause rapid accretion of the disk gas to a dense core
in circular rotation, as is indeed simulated by Fukuda et al. (2000).

Another important finding is that both the NMD and UHC are significantly
warped from the main disk.
The warp of NMD may explain why the H$\alpha$
lobe is highly asymmetric, or lopsided (Veilleux et al 1994).
Warping of the central gas disks appears not a rare case.
In fact, Schinnerer et al. (2000) have shown by their
high-resolution interferometer observations in the CO line,
that the nuclear molecular disk of the Seyfert galaxy NGC 1068 is
significantly warping.
We finlly mention that the origin of the highly inclined nuclear torus
of a few pc scale, as inferred from VLBI jet directions
(Irwin and Seaquist 1988) and maser spots (Sawada-Satoh et al. 2000),
would be somehow related to the warp of its progenitor, NMD and UHC,
from which the torus had accreted.

\v\v
The authors thank the NRO staff for helping us in the observations.
They also thank T. Takamiya for the program to calculate the
surface mass density from rotation curves.

\v\v
\noindent{\bf References}
\v

\r Arimoto, N., Sofue, Y., and Tsujimoto, T. 1996 PASJ 48, 275.

\r  Duric, N., Seaquist, E. R., and Davis, L. E.  1983, ApJL,  273, L11.

\r Duric, N., and Seaquist, E. R. 1988, ApJ, 326, 574.

\r Ford, H. C., Dahari, O., Jakoby, G. H., Crane, P. C., Carduloo, R.
1986 ApJ 311, L7.

\r Fukuda, H., Habe, A., and Wada, K. 2000 ApJ. 529, 109.

\r  Hummel, E., van Gorkom, J. H., and Kotanyi, C. G. 1983, ApJL 267, L5.

\r  Irwin, J. A., and Seaquist, E. R. 1988, ApJ 335, 658.

\r  Irwin, J. A., and Seaquist, E. R. 1991, ApJ  371, 111.

\r  Irwin, J. A. Seaquist, E. R., Taylor, A.R., and Duric, N.
1987, ApJL 313, L91.

\r Irwin, J. A., and Sofue, Y.  1992, ApJ.L. 396, L75.

\r Pietsch, W., Trinchieri, G. and Vogler, A. 1999 AA 340, 351

\r Sawada-Satoh S., Inoue M., Shibata K. M., Kameno S., Migenes V., Nakai N.,
Diamond P.J., 2000, PASJ 52, 421

\r Schinnerer, E., Eckart, A., Tacconi, L. J., Genzel, R., and Downes, D.
2000 ApJ 533, 850.

\r Sofue, Y. 1995 PASJ 47, 527.

\r Sofue, Y. 1996 ApJ 458, 120

\r Sofue, Y. and Irwin, J. A. 1992 PASJ 44, 353

\r Sofue, Y., Tutui, Y., Honma, M., Tomita, A.,  Takamiya, T.,  Koda, J.,
and Takeda, Y. 1999 ApJ. 523, 136.

\r Takamiya, T. and Sofue, Y. 2000, ApJ. 534, 670.

\r Veilleux, S., Cecil, G., Bland-Hathorn, J. et al. 1994 ApJ 433, 48.

\r Young, J. S., Claussen, M. J., and Scoville, N. Z.  1988, ApJ 324, 115.

\newpage

Figure Captions

Fig. 1. Distribution of integrated intensity of the $^{12}$CO ($J=1-0$)
line emission in the central $40'' \times 40''$ area of NGC 3079.
The angular resolution is $1''.62\times1''.34$.
An ultra-high-density molecular core (UHC) is embedded in the nuclear molecular
disk (NMD), which are further surrounded by the main disk component (arrows).
Contours are at 2.5, 5, ...17.5, 20, 25, ..., 35, 40, 50, ... 90 and 100\%
of the peak intensity.

Lower-right inset is a mean-velocity field.
The thick velocity contour near the center is at 1100 \kms, and
the contours are drawn at 50 \kms\ inclement, increasing toward
the south. The thin dashed ellipse outlines the NMD.
Two molecular spiral arms are traced by the thick dashed lines.

Upper-left inset is a $B$-band image from DSS
(vertical extent $8'$). The small white box indicates
the CO map region.

Fig. 2. The central $20''\times20''$ at a higher resolution
of $1''.27\times1''.09$.
The UHC shows up more clearly.
Contours are at 5, 10, 20, ... 90, and 100\% of the peak intensity.

Fig. 3. The upper panel shows
a position-velocity diagram along the major axis of NGC 3079 in CO
with a slit width of 5 arcsec (original resolution $1''.62\times1''.34$).
Contours are at every 5, 10,..., 25, 30, 40, ..,
90 and 100\% of the peak intensity.
The lower panel shows the same, but for the central region
at higher resolution with a slit width 2 arcsec
(original resolution $1''.27\times1''.09$).

Fig. 4. The rotation curve of NGC 3079.

Fig. 5. Surface mass density plotted against radius. The thick curve is
the result for a flat-disk assumption, the thin line for a spherical
assumption.

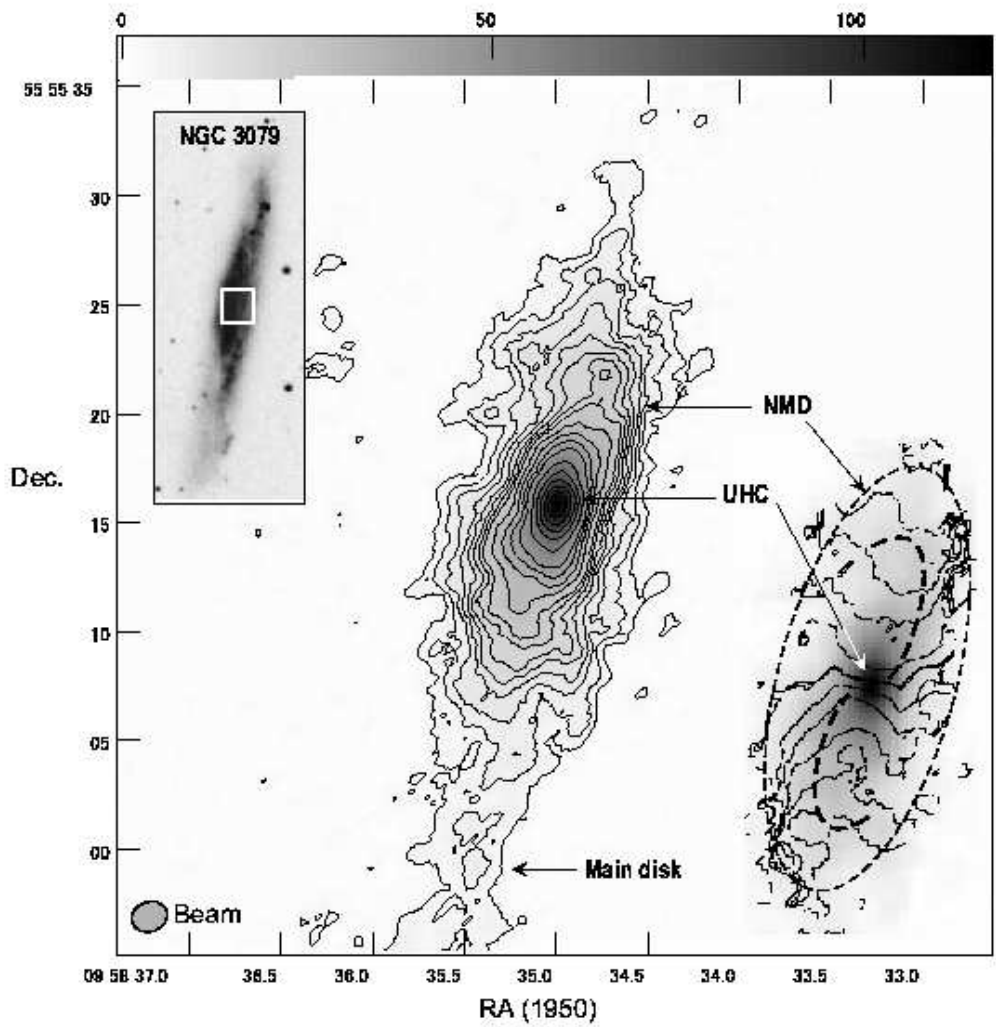
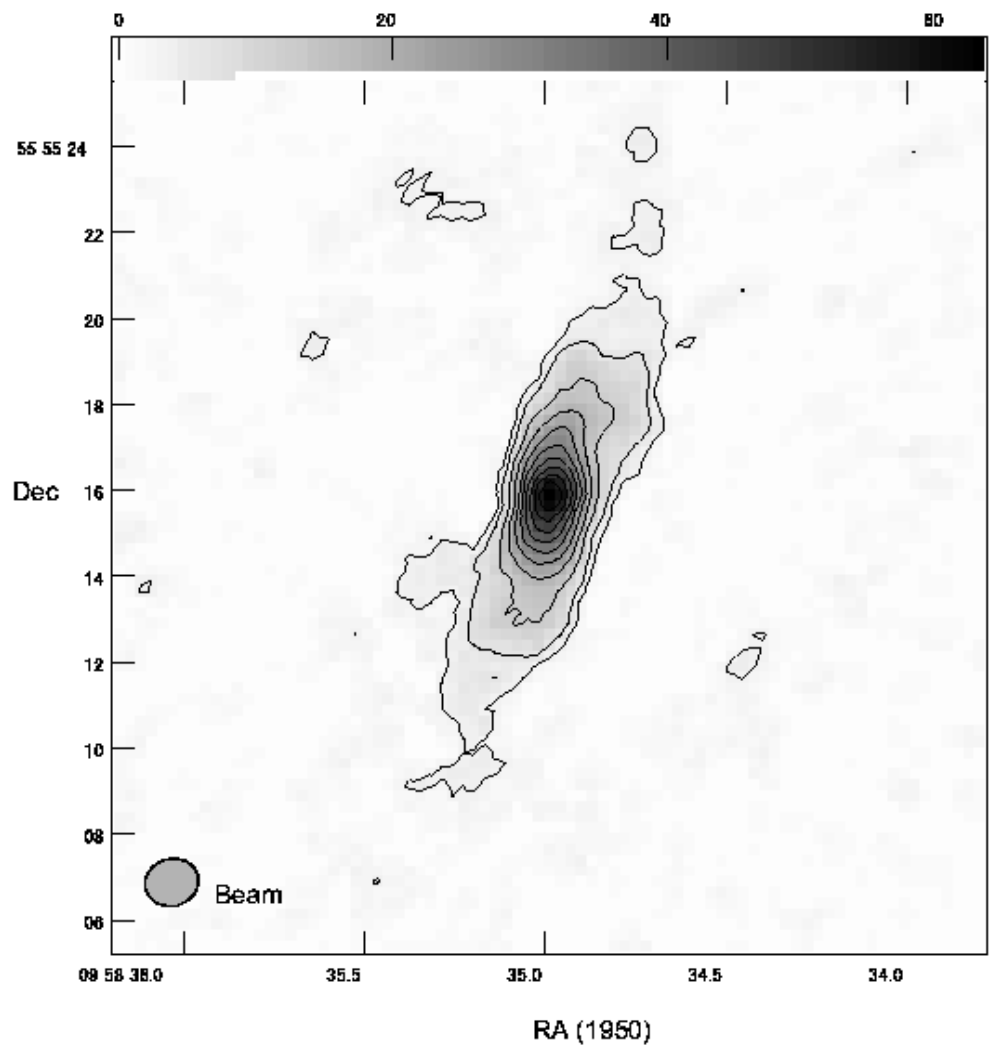
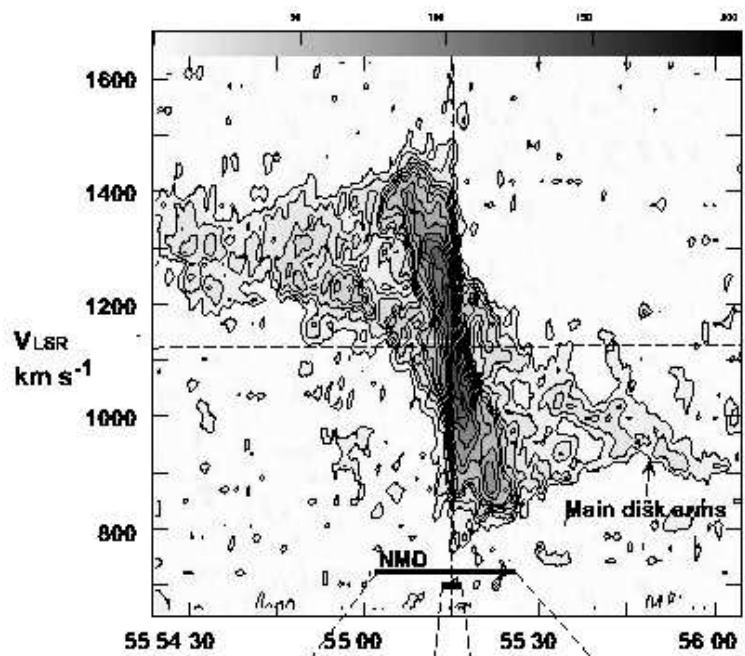
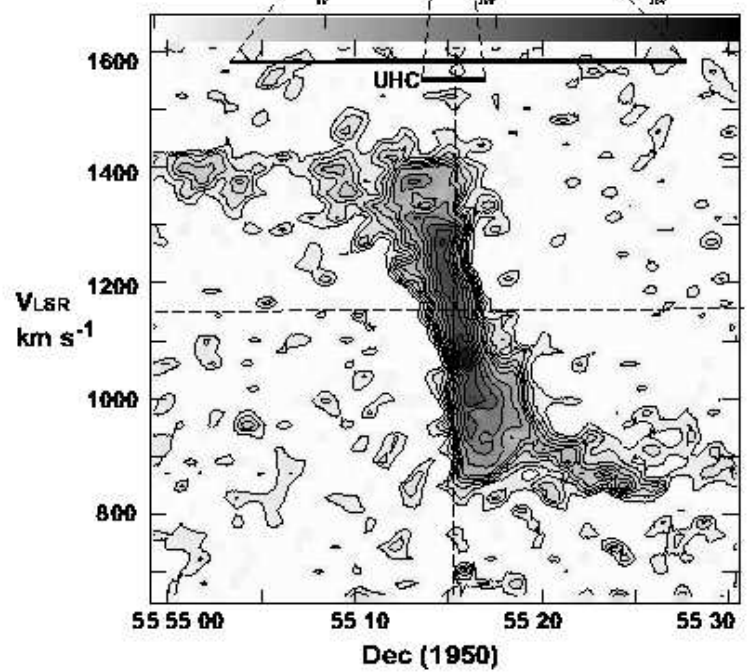
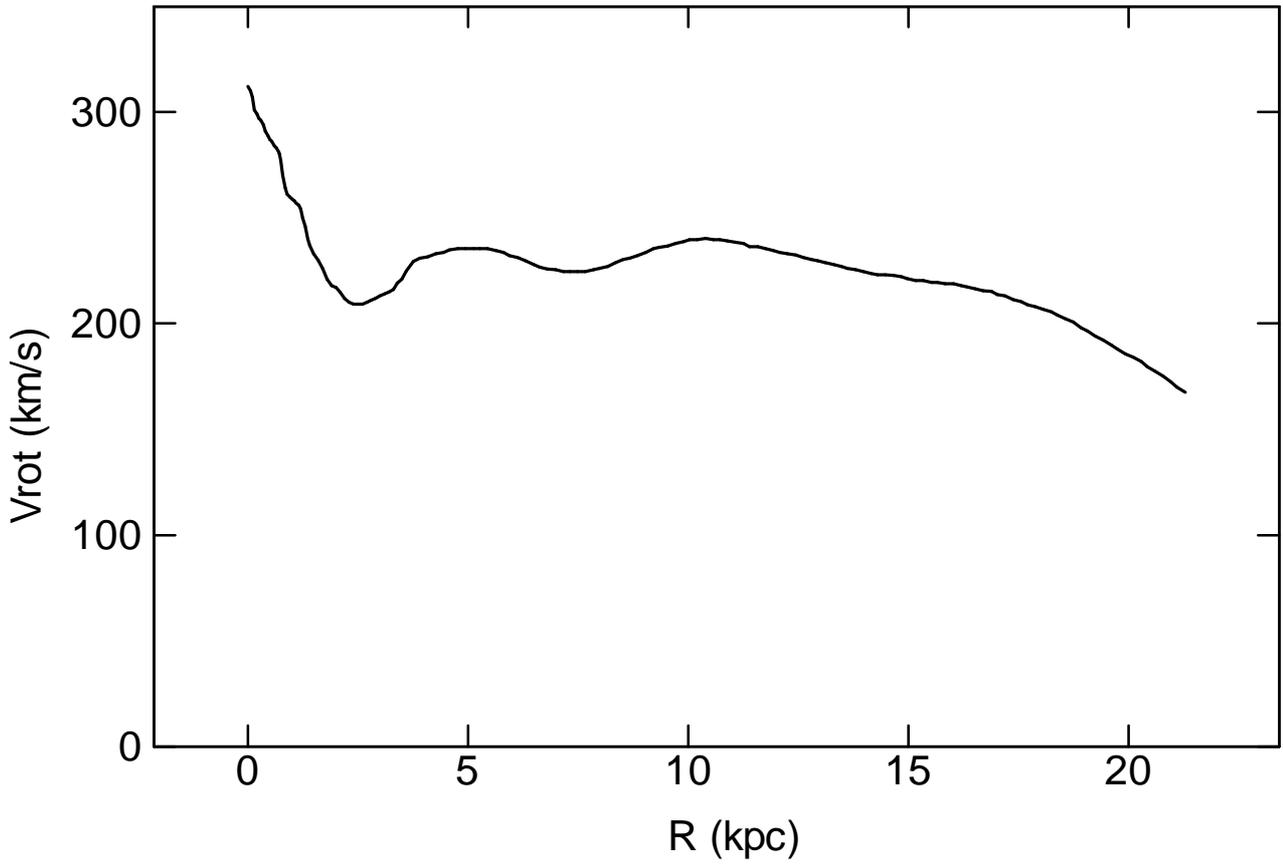
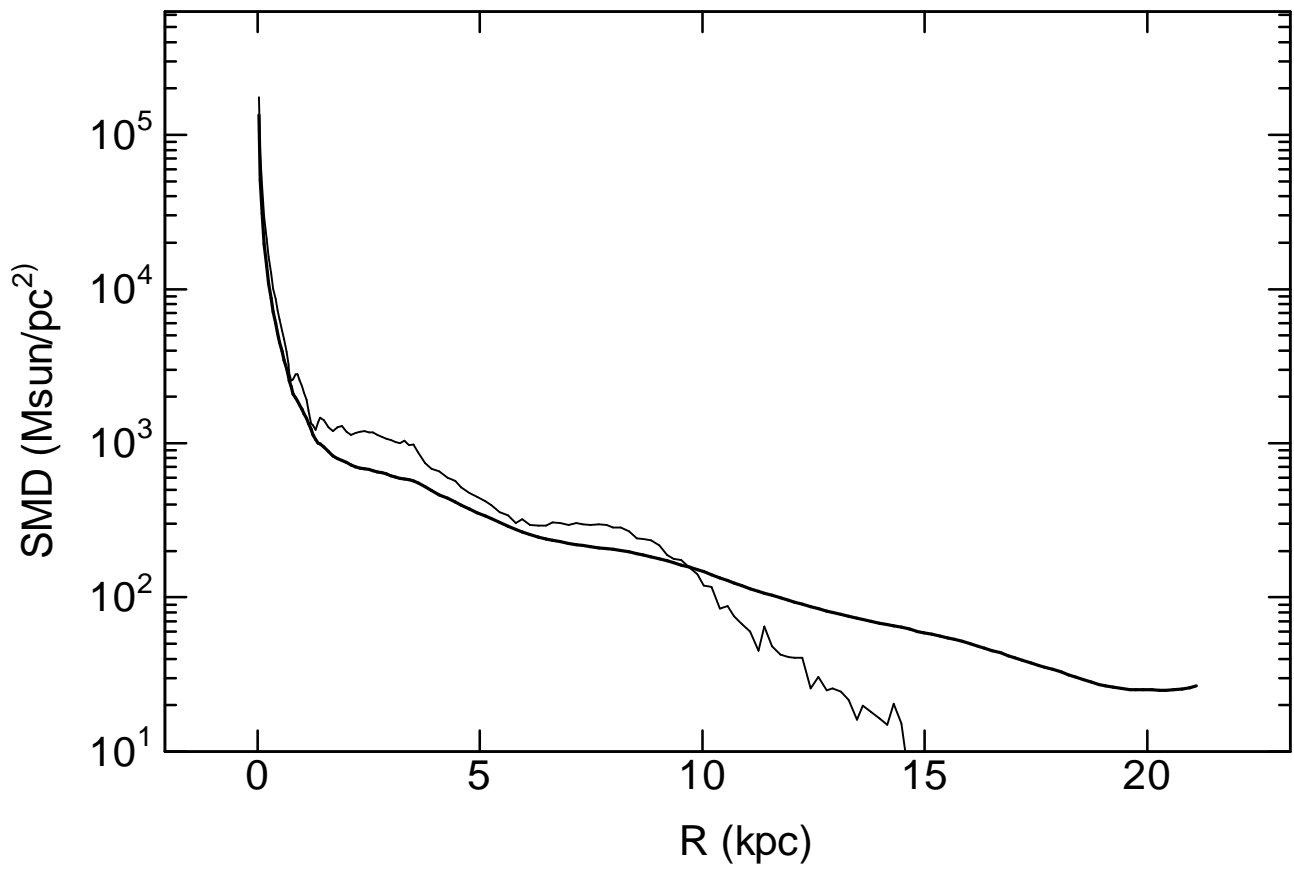
\end{document}